\documentclass{mpe_report}

\usepackage{psfig,graphicx,epsfig}
\usepackage{color}
\usepackage{amsmath,amssymb,epic,eepic,array}

\begin{document}

\pagenumbering{arabic}
\setcounter{page}{212}

 \renewcommand{\FirstPageOfPaper }{212}\renewcommand{\LastPageOfPaper }{215}

\title{A precessing warped accretion disk around the X-ray pulsar Her X-1}
\author{D.~Klochkov\inst{1} \and N.~Shakura\inst{2} \and K.~Postnov\inst{2} \and R.~Staubert\inst{1}
	\and J.~Wilms\inst{1,3}}  
\institute{Institut f\"ur Astronomie und Astrophysik, University of T\"ubingen, Sand 1, 72076 T\"ubingen, Germany
\and Sternberg Astronomical Institute, 119992, Moscow, Russia
\and Department of Physics, University of Warwick, Coventry, CV8 1GA, UK}
\maketitle

\begin{abstract}
We have performed an analysis and interpretation of the X-ray light curve of
the accreting neutron star Her X-1
obtained with the ASM RXTE over the period 1996 February to 2004 September.
The averaged X-ray light curves are constructed by means of adding up
light curves corresponding to different 35 day cycles. A numerical
model is introduced to explain the properties of the averaged light curves.
We argue that a change of the tilt of the accretion disk over the
35\,d period is necessary to account for the observed features and show
that our numerical model can explain such a behavior of the disk and
reproduce the details of the light curve.
\end{abstract}

\section{Introduction}
Her X-1/HZ Her is a close binary system with a 1.7\,d orbital period
containing an accretion powered 1.24\,s X-ray pulsar (\cite{Giacconi73,
Tananbaum72}). The X-ray flux of the source shows a $\sim$35\,d
periodicity which is thought to be due to a counter-orbitally precessing
tilted accretion disk around the neutron star which most likely has a twisted form (\cite{GerBoy76}).

The 35\,d cycle contains two {\em on} states (high X-ray flux) -- 
the {\em main-on} and the {\em short-on} -- separated by $\sim$7-8\,d 
interval of low X-ray flux -- {\em off} state.

According to the generally accepted model
the main-on state starts when the outer disk rim opens the line of sight to the
source. Subsequently, at the end of the main-on state the inner part of the 
disk covers the source from the observer.

An interesting feature of the X-ray light curve are the {\em X-ray dips}
(\cite{Shakura99} and references therein), which
can be separated into three groups: {\em pre-eclipse dips}, which
are observed in the first several orbits after X-ray turn-on, and march from
a position close to the eclipse toward earlier orbital phase in successive
orbits; {\em anomalous dips}, which are observed at 
$\phi_{\rm orb} = 0.45 - 0.65$
and {\em post-eclipse recoveries}, which are occasionally observed as a short
delay (up to a few hours) of the egress from the X-ray eclipse in the first
orbit after turn-on. A general model explaining all types of X-ray dips has been suggested by \cite{Shakura99} where dips are produced by the accretion stream and wobbling outer parts of the accretion disk.

The analysis of Her X-1 observations obtained with the {\em All Sky Monitor} (ASM)
on board of the {\em Rossi X-ray Timing Explorer} (RXTE) satellite (see \cite{Bradt93})
which cover more than 90 35\,d cycles allowed to improve the statistics 
of the averaged X-ray light curve of Her X-1 with respect to previous analysis 
(\cite{Shakura98a,ScoLea99,StillBoyd04}). 
In addition to well-known features 
(pre-eclipse dips and anomalous dips during the first orbit after X-ray 
turn-on) the averaged X-ray light curve shows that anomalous dips and 
post-eclipse recoveries are present for {\em two} successive orbits after the 
turn-on in the short-on state. These details have already been reported by 
\cite{Shakura98a}, but statistics was poor and the authors have not reproduced them with their model.

In this work we substantially improve the numerical model presented by \cite{Shakura99}. Now 
it accounts for the dynamical time scale of the disk and includes explicit calculations of the precession rate at each phase of the 35\,d period. The inclination of the disk is allowed to change during the 35\,d cycle. 

With the improved model we reproduce all details of the averaged light curves including those which were left unexplained previously.

\section{Observations}

For the analysis of the X-ray light curve of Her X-1 we use data from the
ASM (\cite{Levine96}). The archive contains X-ray flux 
measurements in the 2-12 keV band, averaged over $\sim$90 s. The monitoring 
began in February 1996 and continues up to date. The archive is
public and accessible on the Internet 
\footnote{http://xte.mit.edu/asmlc/srcs/herx1.html}.

Preliminary processing of the X-ray light curve was carried out by
the method described by \cite{Shakura98a}.
The goals of this processing are the reduction of the dispersion 
of the flux through rebinning, resulting in a smoothed light curve, and
the determination of the turn-on time of each individual 35\,day
cycle, with the aim to classify the cycles into two classes: turn-on 
around orbital phase $\sim 0.2$, and turn-on around orbital phase 
$\sim 0.7$ (they appear with about equal probability).

Using the RXTE/ASM (with $\sim$5\,cts/s from Her~X-1 in the main-on)
details cannot be explored in individual 35 day cycles.
However, if we construct averaged light curves through superposition of many 35\,d light curves
common details (e.g. X-ray dips, post-eclipse recoveries) become recognizable.

This has been done e.g. by \cite{Shakura98a,ScoLea99,StillBoyd04}. The ASM
archive has grown considerably, allowing to construct averaged light
curves with smaller dispersion than in previous works.

\begin{figure}

\centerline{\psfig{file=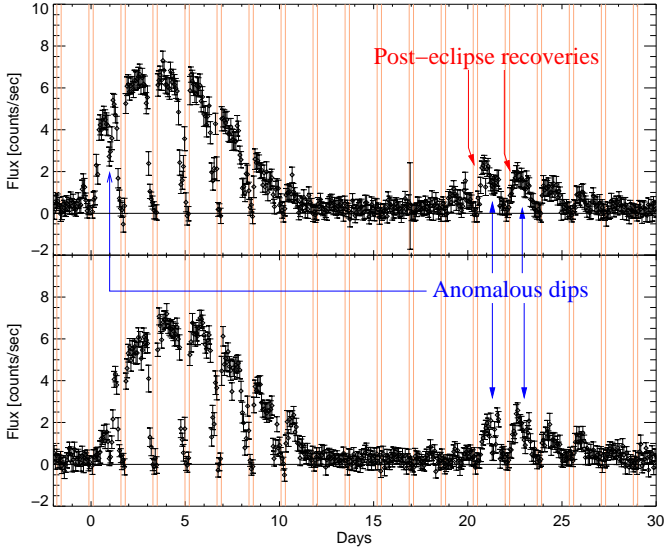,width=8.8cm,clip=} }

\caption{Averaged X-ray light curves of \hbox{Her X-1} 
corresponding to cycles with turn-ons near orbital phase
$\sim$0.2 (top) and $\sim$0.7 (bottom). Vertical lines show X-ray eclipses.
\label{aver}}
\end{figure}

As mentioned above, in most cases turn-ons occur near orbital phases 0.2 
and 0.7. Thus all 35\,d cycles have been divided
into two groups -- with the turn-on near $\phi_{\rm orb} = 0.2$ and
near $\phi_{\rm orb} = 0.7$. Inside each group the light curves were 
superposed and averaged, after shifting them in such a way that the
eclipses coincided.
The superposed light curves are shown in Fig.~\ref{aver}.

\begin{figure}
\centerline{\psfig{file=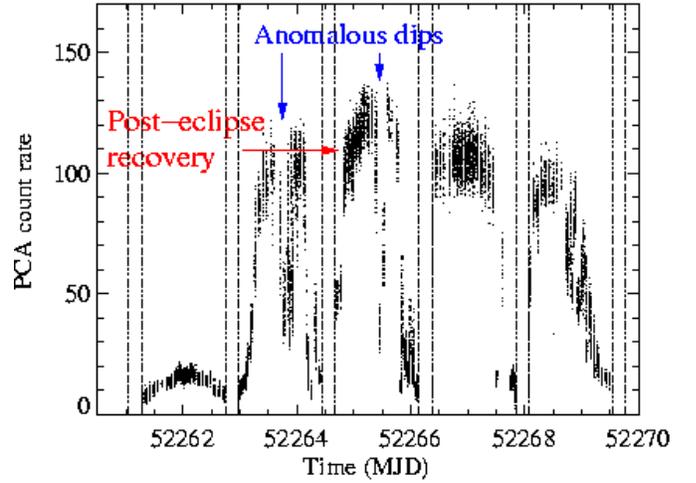,width=8.8cm,clip=} }
\caption{3-20 keV PCA light curve of \hbox{Her X-1} during a 
short-on state.
\label{shorton}}
\end{figure}

In the short-on state complicated dip patterns can be observed.
Anomalous dips near orbital phase $\phi_{\rm orb} = 0.5$
and post-eclipse recoveries are present on two successive orbits after the 
beginning of the short-on state. This can also be seen in one individual
short-on observed by RXTE/PCA shown in Fig.~\ref{shorton}
(see also \cite{Leahy00, Oosterbroek00, InamBaykal05}).

\section{The model}

As mentioned in the Introduction, it is believed that the alternation of the {\em On} and {\em Off} states is caused through shadowing by a counter-orbitally precessing tilted twisted accretion disk.
In our model (\cite{Shakura99}) pre-eclipse dips occur when the 
accretion stream crosses the observer's line of sight before entering the 
disk. This can happen only if the stream is non-coplanar to the systems orbital
plane. The reason for the stream to move out of the orbital plane 
is non-uniform X-ray heating of the optical stars atmosphere by the X-ray 
source, which produces a temperature gradient near the inner Lagrange point. 
The non-uniformity of the heating comes from the partial shadowing of the 
optical star surface by the accretion disk.
Furthermore, such a stream forces the outer parts of the 
accretion disk to be tilted with respect to the orbital plane. The tidal 
torques cause the disk to precess in the direction opposite to the orbital 
motion. Due to tidal torques and the dynamical action of the accretion stream 
the outer parts of the disk develop a notable wobbling (nutational) motion 
twice the synodal period. For an observer, the X-ray source can be screened 
by the outer parts of the disk for some time during the first orbit after the 
X-ray turn-on. This causes anomalous dips and post-eclipse recoveries.

\subsection{Numerical calculations of the disk motion}

The method of calculation of the disk motion under the action of tidal forces and the accretion stream is described by \cite{Shakura99}. The disk was approximated by a solid ring with the radius equal to the outer radius of the disk. The inclination of the disk with respect to the orbital plane was assumed to be constant. Under these assumptions the dynamical time scale of the disk $t_{\rm d}=M_{\rm d}/\dot M$ (where $M_{\rm d}$ and $\dot M$ are the mass of the disk and the accretion rate correspondingly) which characterizes the dynamical action of the stream turned out to be significantly shorter than the viscous time scale which leads to inconsistency. Also this model failed to explain some details of the averaged light curve, namely, anomalous dips and post-eclipse recoveries on two orbits after the secondary turn-on.

In this work the model is substantially improved.
In order to reconcile the dynamical and viscous time scales of the disk we introduce two characteristic radii: the outer radius of the disk $r_{\rm out}\sim 0.3a$ which determines 
the tidal wobbling amplitude, and the effective radius $r_{\rm eff}\sim 0.18a$ 
which determines the mean precession motion of the disk. The value of this effective radius is found from the requirement that the net precession period of the entire disk be equal to the observed 
value 20.5$P_{\rm orb}$ and the dynamical time scale $t_{\rm d}=M_{\rm d}/\dot M$ 
(which characterizes the dynamical action of the stream) be 10 days. 
This value for the dynamical time is chosen to be of order of the viscous 
time from the impact radius $\sim$0.1$a$ which is around 10 days and scales as $r^{-3/2}$.

The inclination of the disk is allowed to change with the precessional phase. It requires that the rate of precession was calculated for each precessional phase separately since it depends on the disk inclination. 

It is important to note that the region of the disk beyond 
$r_{\rm imp}$, where is no matter supply (called the ``the stagnation zone'') 
and which mediates the angular momentum transfer outwards for accretion to 
proceed, can react to perturbations induced by the stream impact much faster 
than on the viscous time scale. Indeed, in a binary system, tidal-induced 
standing structures can appear in the outer zone of the accretion disk 
(see e.g. \cite{Blondin00} and references therein) and perturbations 
of angular momentum can propagate through this region with a velocity close 
to the sound speed (while the matter will accrete on the much slower viscous 
time scale!). Recent analysis of broad-band variability of SS 433 (\cite{Revnivtsev05}) 
suggests that such a picture is realized in the accretion disk in that source.
So the entire disk (inside and beyond $r_{\rm imp}$) reacts to changes in the mass transfer rate through the accretion stream on a time scale not longer than the viscous time scale of the disk from the impact point (i. e. $\sim 10$ days).

\subsection{Evidence for a change in the tilt of the disk}

The fact that during the short-on state of Her X-1 more absorption dips 
appear on the averaged X-ray light curve  (the anomalous dips and 
post-eclipse recoveries during two successive orbits of the 
averaged short-on) as well as in an individual short-on observed by PCA
(Fig.~\ref{shorton} and \cite{InamBaykal05}) suggests that the  
angle $\epsilon$ between the disk and the line of sight remains close to zero 
during the first two orbits after the turn-on in the short-on state. This is 
in contrast to the main-on state, where such features are observed only during
one orbit after turn-on. We failed to reproduce the observed behavior assuming constant inclination of the disk. The observations imply a generic asymmetry between the beginning of the main-on and short-on states. This asymmetry can take place only if the disk tilt $\theta$ changes with precessional phase.

The physical reason for this could be the periodic change of the mass transfer rate from the inner Lagrange point into the disk. If free precession of the neutron star is ultimately 
responsible for the 35-day cycle in Her X-1 (\cite{Brecher72,Truemper86,Shakura95,Shakura98b}), 
the conditions of  X-ray illumination of the optical stars atmosphere will periodically change with 
precession phase. This in turn will lead to changes in the velocity 
components of the gas stream in the vicinity of the inner Lagrangian point and
hence in the matter supply rate to the accretion disk.
Recently, X-ray pulse profiles evolution with the 35\,d 
phase was successfully reproduced both in the main-on and short-on
states in the model of freely precessing neutron star with complex 
surface magnetic field structure (\cite{Ketsaris00,Wilms03}).

\section{Results}

The physical picture of the change of the disks tilt adopted here is as follows. 
The rate of mass transfer $\dot M$ supplied by HZ Her (which 
for a given moment can be different from the  
accretion rate of the neutron star because the viscous time scale for mass
transport through the disk both delays and smoothes the mass flow)
changes periodically over 
the precession cycle in response to changing illumination of 
the optical star atmosphere  due to free precession of the neutron star. 
The streams action causes the outer disk to precess slower than it would 
do if only tidal torque was acting, and it also changes the outer 
disk's tilt on the dynamical time scale of about 10 days (see above), which 
is sufficient to explain periodic disk tilt variations over the precession 
cycle. The wobbling of the outer disk is mainly due to tidal forces 
(the dynamical action of the stream provides minor contribution). 
The viscous time scale in the ``stagnation zone'' is much longer, so the 
viscous torques cannot smooth out the external disk variations.

\begin{figure}
\centerline{\psfig{file=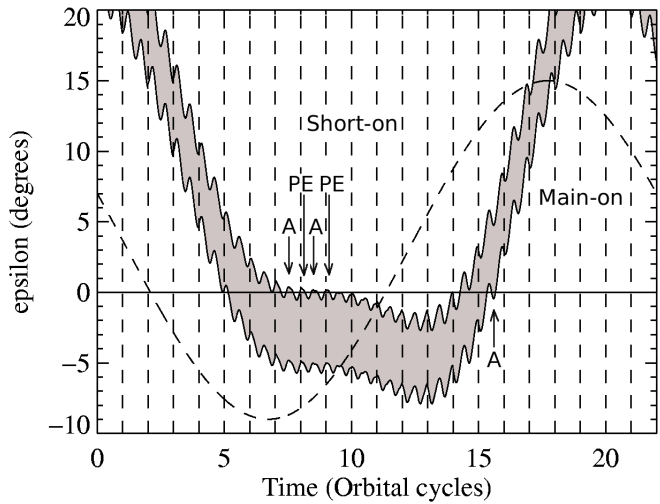,width=8.8cm,clip=} }
\caption{The angle $\epsilon$ between the direction from the 
center of the neutron star to the observer and to the outer parts of the 
accretion disk calculated using our model. The sine-like dashed
curve shows schematically the inner accretion disk region.
The source is screened by the disk when the observer is in the dark area or
  between this area and the inner disk line. 
Arrows mark anomalous dips (A) and post-eclipse recoveries (PE).
\label{epsilon}}
\end{figure}

Figure~\ref{epsilon} shows the result of modeling of the disk motion. 
The angle $\epsilon$ (vertical axis) is the angle between the direction from the center of
the neutron star to the observer and to the outer parts of the accretion 
disk with non-zero thickness.
The complex shape of the disks wobbling is clearly seen. The sine-like dashed
curve shows schematically the same angle for the inner accretion 
disk regions which eclipse the X-ray source at the end of on-states. 
The horizontal line is the observer's plane and the vertical
dashed lines mark centers of the binary eclipses. The main-on and
short-on states are indicated. 
The source is screened by the disk when the observer is in the dark area
or between this area and the inner disk line. 
It is seen that the wobbling
effects can be responsible for the observed (several) anomalous
dips (marked with ''A'') and post-eclipse recoveries at the beginning of the short-on
state (marked with ''PE'').

\section{Conclusions}

The following main results have been obtained from analyzing
and modeling RXTE/ASM X-ray light curves of Her~X-1.
\begin{enumerate}
\item The shape of the averaged X-ray light curves is determined more
accurately than was possible previously (e. g. by \cite{Shakura98a}).
\item We have significantly improved the model developed in \cite{Shakura99} by including the calculation of the dynamical time scale of the disk, more accurate calculation of its precessional motion and allowing the disk to change its inclination with respect to the orbital plane during the 35\,d cycle.
\item With the improved model we successively reproduced observed details of the X-ray light curve including the anomalous dips in two successive orbits after the beginning
of the short-on state and the post-eclipse recovery after the first eclipse which were left unexplained previously.
\item We argue that the changing of the tilt of the accretion disk with 
35\,d phase is necessary to account for the appearance of these features and
the observed duration of the main-on and short-on states.
\end{enumerate}

\begin{acknowledgements}
In this research we used data obtained through the High Energy Astrophysics
Science Archive Research Center Online Service, provided by the NASA/Goddard
Space Flight Center.
We thank A. Santangelo for useful discussion.
The work was supported by the DFG grant Sta 173/31-2 and
436 RUS 113/717/0-1 and the corresponding RBFR grant RFFI-NNIO-03-02-04003.
KP also acknowledges partial support through RFBR grant 03-02-16110.

We gratefully acknowledge the support by the WE-Heraeus foundation.
\end{acknowledgements}



        \clearpage

\end{document}